\documentclass[]{aastex}

\usepackage{emulateapj5}
\usepackage{onecolfloat}
\usepackage{graphicx} 
\usepackage{fancyheadings} 
\usepackage{ulem}
\usepackage{rotating}
\usepackage{lscape}

\newcommand{\etal}{et al.\ }

\newcommand{\lya}{Ly$\alpha$ }
\newcommand{\kms}{km~s$^{-1}$ }

\newcommand{\perd}{\;\;\; .}
\newcommand{\cmma}{\;\;\; ,}

\newcommand{\dv}{$\Delta v$}
\newcommand{\dvm}{$\delta v$}
\newcommand{\dsiiv}{$\Delta v_{\rm Si IV}$}
\newcommand{\dciv}{$\Delta v_{\rm C IV}$}
\newcommand{\dlow}{$\Delta v_{\rm low}$}
\newcommand{\daliii}{$\Delta v_{\rm Al III}$}
\newcommand{\sigdel}{$\sigma_{\delta v}$}
\newcommand{\eslash}{$\rm \acute e$}

\begin{document}

\twocolumn[%
\submitted{accepted by the Astrophysical Journal August 5, 2000}

\title{IONIZED GAS IN DAMPED \lya\ PROTOGALAXIES: I. MODEL-INDEPENDENT
INFERENCES FROM KINEMATIC DATA }

\author{ ARTHUR M. WOLFE\altaffilmark{1}\\
Department of Physics, and Center for Astrophysics and Space Sciences; \\
University of California, San
Diego; \\
C--0424; La Jolla; CA 92093\\
{\bf awolfe@ucsd.edu}}

\author{and}

\author{ JASON X. PROCHASKA\altaffilmark{1}\\
The Observatories of the Carnegie Institute of Washington; \\
813 Santa Barbara St.\\
Pasadena, CA; 91101\\
{\bf xavier@ociw.edu}}

\begin{abstract}

We investigate the kinematics of ionized and neutral gas
in a sample of 35 damped {\lya} systems
(protogalaxies) using accurate quasar spectra
obtained with HIRES,  the Echelle spectrograph
on the Keck I 10 m telescope. Velocity profiles with resolution of $\sim$ 8
{\kms} are obtained for high ions such as C IV   and Si IV,
and for intermediate ions such as Al III. Combining these profiles  with
similar quality profiles obtained previously for low ions such as Fe II,
we investigate the kinematic state of damped {\lya} protogalaxies in
the redshift range 1.8 $< \ z \ <$ 4.4 by comparisons between data for
various ion pairs.

We find the damped {\lya} protogalaxies comprise distinct kinematic subsystems:
a low ion subsystem in which the low ions are physically associated with
intermediate ions, and a high ion subsystem containing neither
low nor intermediate ions.
The evidence for two subsystems stems from (a)
differences between the widths of the  velocity profiles, (b)
misalignment in velocity space between the narrow components that
make up the profiles
in each subsystem, and (c) significant dissimilarities between
the mean velocities of the high ion and low ion velocity profiles. In every case
we find that test statistics
such as velocity width and various asymmetry parameters
distribute differently
for low ions than for high ions.
We also find
the absence of
intermediate and low ions from the high-ion subsystem to indicate the
latter is optically thin at the Lyman limit.

Despite misalignment between their velocity components, the low and high
ion kinematic subsystems are interrelated. This is indicated by detection
of a statistically significant C IV versus low ion cross correlation function.
It is also indicated by a systematic effect where the C IV velocity
widths are greater than or equal to the low ion velocity widths
in 29 out of 32 systems. These phenomena are consistent with the location of
the two
subsystems in the same potential well.

 \end{abstract}

]
\altaffiltext{1}{Visiting Astronomer, W.M. Keck Telescope.
The Keck Observatory is a joint facility of the University
of California and the California Institute of Technology.}

\pagestyle{fancyplain}
\lhead[\fancyplain{}{\thepage}]{\fancyplain{}{WOLFE \& PROCHASKA}}
\rhead[\fancyplain{}{IONIZED GAS IN DAMPED {\lya} PROTOGALAXIES}]{\fancyplain{}{\thepage}}
\setlength{\headrulewidth=0pt}
\cfoot{}

\section{INTRODUCTION}

For the past several years we have been studying the kinematics
of gas in damped {\lya} systems.
We have focused on the {\em neutral}
gas (Prochaska \& Wolfe 1997; 1998: hereafter PW1, PW2)
because of evidence suggesting this to be the source of baryons
for stars in current galaxies (Wolfe 1995, 1997; Kauffmann 1996).
We
used the HIRES Echelle spectrograph \citep{vog94}
on the Keck I 10 m telescope to obtain accurate
velocity profiles of low ions such as Fe II, Si II,
Ni II, and Al II as they
trace the kinematics of the neutral gas.
The evidence for this is
the large Al II/Al III ratios detected in
most damped {\lya} systems, which
indicate
the singly ionized  species to be associated with
neutral rather than ionized gas, provided the ionizing radiation
is supplied by external sources \citep{pro96}.

The low-ion velocity profiles that we have measured
comprise multiple narrow components that
are not randomly distributed in velocity space. Rather the strongest
component tends to occur at the profile edge.
In PW1 and PW2 we used these properties
to test various models for damped {\lya} systems.
We adopted
Monte Carlo techniques
by sending random sightlines through gaseous configurations
specified by the cosmological model, the geometry and physics of the configuration, etc.
We then computed distributions of test statistics such as the
profile line widths and compared them with the empirical distributions.
We first tested a simple model
in which dark-matter halos
halos enclose {\em identical} randomly oriented exponential disks with
rotation speeds $V_{rot}$ $\approx$ 250 {\kms} and exponential scale-height
$h$ $\approx$ 0.3$R_{d}$, where $R_{d}$ is the radial scale-length
(PW1). We found this model to satisfy
all the statistical tests.
Its principal disadvantage, however, is
it is not set
in a cosmological context.
We then tested semi-analytic versions of standard adiabatic CDM cosmologies (SCDM) in which
the neutral gas is confined to centrifugally supported disks in dark matter
halos drawn from a computed mass function.
The predominance of objects with low-mass and low $V_{rot}$, results
in line widths that are too low to match the observed distribution which extends
to 300 {\kms}. In particular, the SCDM models considered by
\cite{kau96} were found to be incompatible with the kinematic data.
Furthermore, \cite{jedpro98} showed that no CDM cosmology is
consistent with models assuming the damped {\lya} systems to be
single centrifugally supported disks within dark-matter halos.
On the other hand \cite{haeh98}
numerically simulated SCDM models with gas and dark matter, and found
that at $z$ $\sim$ 2
the damped {\lya} gas was distributed in low-mass protogalactic clumps rather than centrifugally
supported disks. The combination of infall and chaotic motions apparently
reproduce the kinematic data.

This paper focuses on the kinematics of the {\em ionized} gas.
Our aim is to place further constraints on galaxy formation models
by studying gas located
outside the neutral zones giving rise to
damped {\lya} lines. Ionized gas is  a generic feature
of such models because initially the baryons are heated and ionized
as they virialize in
the potential wells of dark-matter protogalactic halos.
However, the  indicated velocity field is not unique,
with some models predicting radial collapse to neutral disks (Mo \& Miralda-Escud{\eslash} 1996)
,
while others envision chaotic motions of ionized and neutral blobs (Haehnelt {\etal} 1997).
As a result, determining the
velocity
structure of the ionized gas should help to
clarify
crucial events  in the galaxy formation process
(see accompanying paper by Wolfe \& Prochaska
2000; hereafter paper II).

In order to trace the velocity structure of the ionized gas
we used HIRES to obtain accurate velocity profiles of the
C IV, Si IV, and Al III ions; these are shown in $\S$ 2.
Throughout this paper we assume the C IV ion to represent
the highly ionized gas, since we have a larger number of accurate C IV than Si IV
velocity profiles.
In $\S$ 3 we construct
frequency distributions of the profile line width, {\dv},
for each ion. We compare these with each other and with
the {\dv} distribution for low ions such as Fe II or Si II
which presumably trace the kinematics of the neutral gas.
To further intercompare the low ion and high ion gas we examine
correlations between the kinematics of low ions and high ions.
To that end
we consider
the difference between the means
of low ion and C IV profiles.
We also consider the ratios of the {\dv}'s for various ion
pairs.
We then cross-correlate the velocity profiles of
various ionic pairs.
In $\S$ 4 we present model-independent conclusions following
from our results.

\begin{table}[ht] \footnotesize
\begin{center}
\caption{{\sc IONIC TRANSITIONS IN DAMPED {\lya} SYSTEMS} \label{tab:obs}}
\begin{tabular}{llccccc}
&&\multicolumn{4}{c}{Transition}&\\
\cline{3-6}
QSO & $z_{abs}$
& C IV
& Si IV
& Al III & Low &Ref.  \\
\tableline
Q0100$+$1304 & 2.309 & 1548& 1393 & 1854 & Ni II 1741  &W\tablenotemark{a}\cr
Q0149$+$33 & 2.14075 & 1548& 1393 & 1854 & Fe II 1608&W \cr
Q0201$+$3634 & 2.4628 & 1550 & 1393 & 1862 & Si II 1808  &W\cr
Q0216+0803\tablenotemark{b} & 2.2930 & 1550& --- & 1862 & Si II 1808&S\tablenotemark{c} \cr
Q0347$-$3800 & 3.0247 & 1548& 1393 & --- & Fe II 1608&W \cr
Q0458$-$020  &2.03955 &---   &---   & 1854 & Cr II 2056  &W\cr
Q0528$-$2505A\tablenotemark{d} & 2.14104 & 1550& --- & --- & Si II 1808 &S\cr
Q0841$-$0203A & 2.374518 & 1548& --- & 1854 & Ni II 1741  &W\cr
Q0841$-$0203B & 2.476219 & 1548& 1393 & 1854 & Fe II 1608  &W\cr
Q0930$+$28   & 3.23525 & 1548 & 1393 & --- & Fe II 1608&S\cr
Q0951$-$04A & 3.85669 & --- & 1393 & --- & Si II 1526&W\cr
Q1055+46 & 3.3172   & 1548 & ---- & --- & Fe II 1608 & S \\
Q1104$-$18   & 1.661375 & 1550& --- & 1854 & Si II 1808&S \cr
Q1108-07 & 3.607619 & 1548 & 1402 & --- & Fe II 1608 & W \\
Q1202$-$0725 & 4.38285 & 1548& 1393 & --- & Si II 1304 &S\cr
Q1215+3322 & 1.9991 & 1548 & 1393 & 1854 & Si II 1808  &W\cr
Q1223$+$17&2.466083 & 1550& 1402 & 1862  &Si II 1808  &S\cr
Q1331+1704\tablenotemark{a} & 1.77636 & 1550& --- & 1854 & Si II 1808  &W\cr
Q1425+6039 & 2.8268 & 1550& 1402 & --- & Fe II 1608 &S\cr
Q1759$+$7500 & 2.6253 & 1550& --- & 1854 & Si II 1808&W \cr
Q1850$+$40   & 1.99016 & ---& --- & 1862 & Zn II 2026&S \cr
Q1946+7658A & 1.7382 & 1548& --- & 1854 & Si II 1808 & S\cr
Q1946+7658B & 2.8443 & 1550& 1402 & --- & Si II 1304 & S\cr
Q2206$-$1958A\tablenotemark{a} & 1.920 & 1548& --- & 1854 & Ni II 1741  &W\cr
Q2206$-$1958B & 2.07623 & 1550& 1402 & --- & Al II 1670  &W\cr
Q2212$-$1626 & 3.6617 & 1548& 1402 & --- & Si II 1304&S \cr
Q2230$+$02   & 1.858536 & 1550& 1402 & 1862 & Si II 1808&W \cr
Q2231$-$0015 & 2.06615 & 1548& 1393 & 1854 & Si II 1808  &W$+$S\cr
Q2233$+$13   & 3.14927 & 1548 & 1402 & --- & Fe II 1608&S \cr
Q2237$-$0608 & 4.0803 & 1548& 1393 & --- & Al II 1670&S \cr
Q2343$+$12   & 2.42969 & 1550& 1393 & 1854 & Si II 1808&S \cr
Q2344$+$12   & 2.537789 & 1548& 1402 & --- & Al II 1670&S \cr
Q2348$-$1400\tablenotemark{a} & 2.2794 & 1548& 1393 & 1862 & Fe II 1608&W \cr
Q2359$-$0203A\tablenotemark{a} &2.095067 & 1550& --- & 1854 & Si II 1808  &W\cr
Q2359$-$0203B & 2.153934 & 1550& --- & 1862 & Si II 1526  &W\cr
\tableline
\end{tabular}
\end{center}
\tablenotetext{a}{Data collected by our group}
\tablenotetext{b}{C IV profiles exhibit some saturation}
\tablenotetext{c}{Data kindly provided by W. L. W. Sargent and collaborators}
\tablenotetext{d}{The $z=2.811$ damped system toward Q0528$-$2505
was omitted because its redshift exceeds that of the background QSO to which
it may be associated.}
\end{table}

\section{DATA: OBSERVED VELOCITY PROFILES}

Table 1 lists the sample of damped {\lya} systems for which we have obtained
velocity profiles. The spectra were acquired with HIRES at a resolution
with FWHM $\approx$ 6$-$8 {\kms} and were reduced according to
procedures outlined in PW1 and PW2.
The coordinate name of the background QSO is in column 1.
Column 2 specifies the absorption redshift of the damped system.
The entries in columns 3$-$5 contain the specific C IV, Si IV,
and Al III transitions used in our analysis.
Entries with horizontal lines indicate that statistically
significant profiles were not observed. Column 6 lists the low ion transitions
used in our analysis and
Column 7 gives the data references. Here and throughout the paper, the term
low ion refers to ions such as Fe II and Si II, the dominant ionization
states in neutral gas with  large optical depths at the Lyman limit , i.e., ${\tau_{LL}}$
$>$ 10$^{4}$. The ionization potential (IP) of such ions is greater than 1 Ryd, i.e. 13.6 eV,
while the IP of the next lower state (typically  the neutral atom except
for O) is
less than 1 Ryd.
The term high ion refers to ions where the next lower state has
IP~$\gg$~1~Ryd, for example Si IV and C IV.
Photoionization studies show these ions to be in gas which
is optically thin at the Lyman limit or in gas with moderate
Lyman-limit optical depths (${\tau_{LL}}$ $<$ 10$^{2}$).
Finally, we introduce the term intermediate ions to describe those ions
where the next lower ionization state has IP $\gtrsim$~1~Ryd (e.g.\
Al III).
The same photoionization studies show that at column densities
required for detection (i.e. above 10$^{14}$cm$^{-2}$)
these ions arise in gas which is optically thick at
the Lyman limit. As a result, association of high ions
with intermediate ions indicates the gas is optically thick at the Lyman limit.

\begin{figure*}[ht]
\begin{center}
\includegraphics[height=8.0in, width=6.3in]{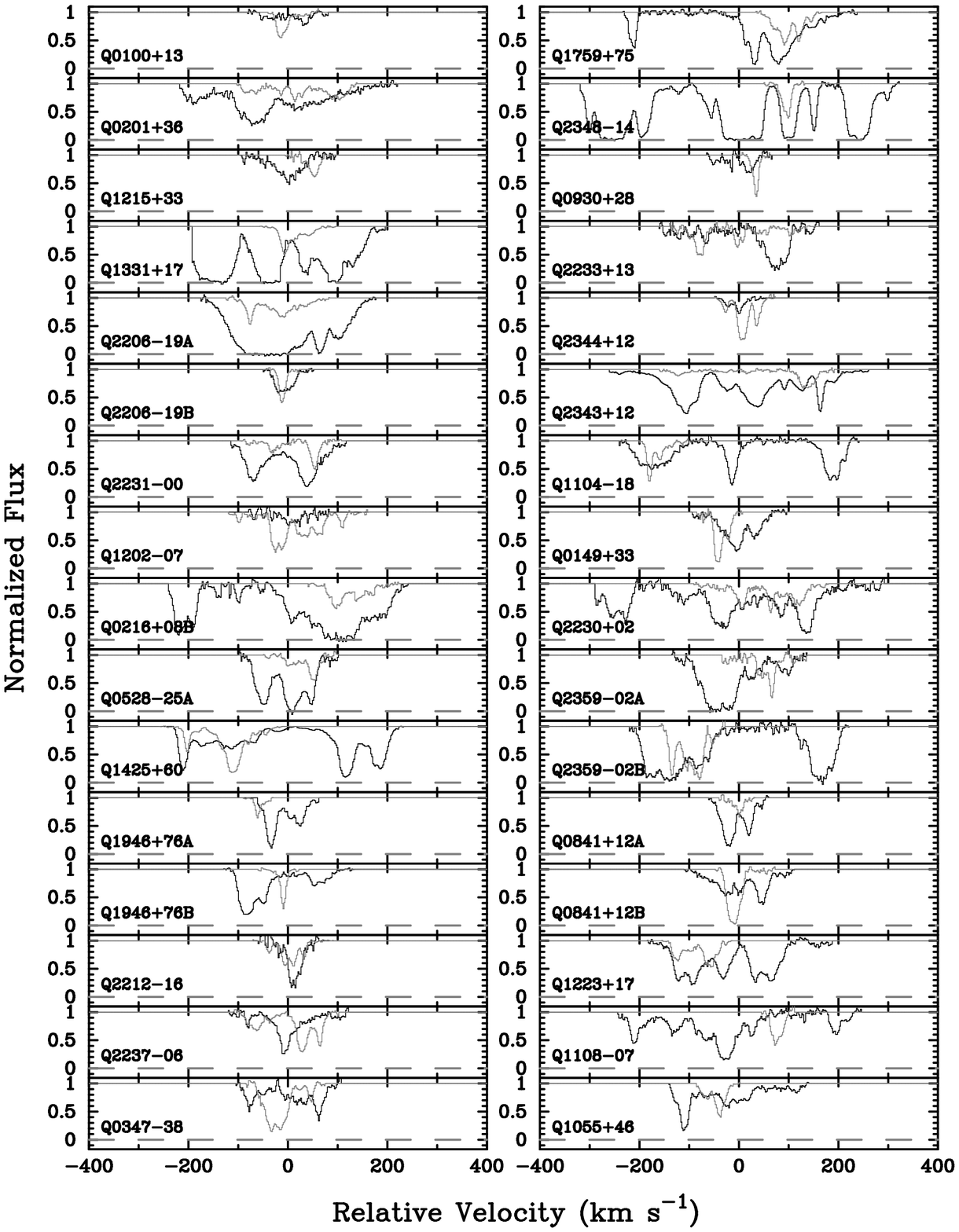}
\end{center}
\caption{Velocity
profiles of C IV (dark lines)  and low ion (grey lines)
transitions for all damped {\lya} protogalaxies in Table1
in which HIRES C IV and low ion profiles have been measured. The velocity
$v$ = 0 corresponds the absorption redshift, $z_{abs}$, shown in
the Table. }
\label{CIVvsLow}
\end{figure*}

\begin{figure*}[ht]
\includegraphics[height=5.8in, width=6.2in]{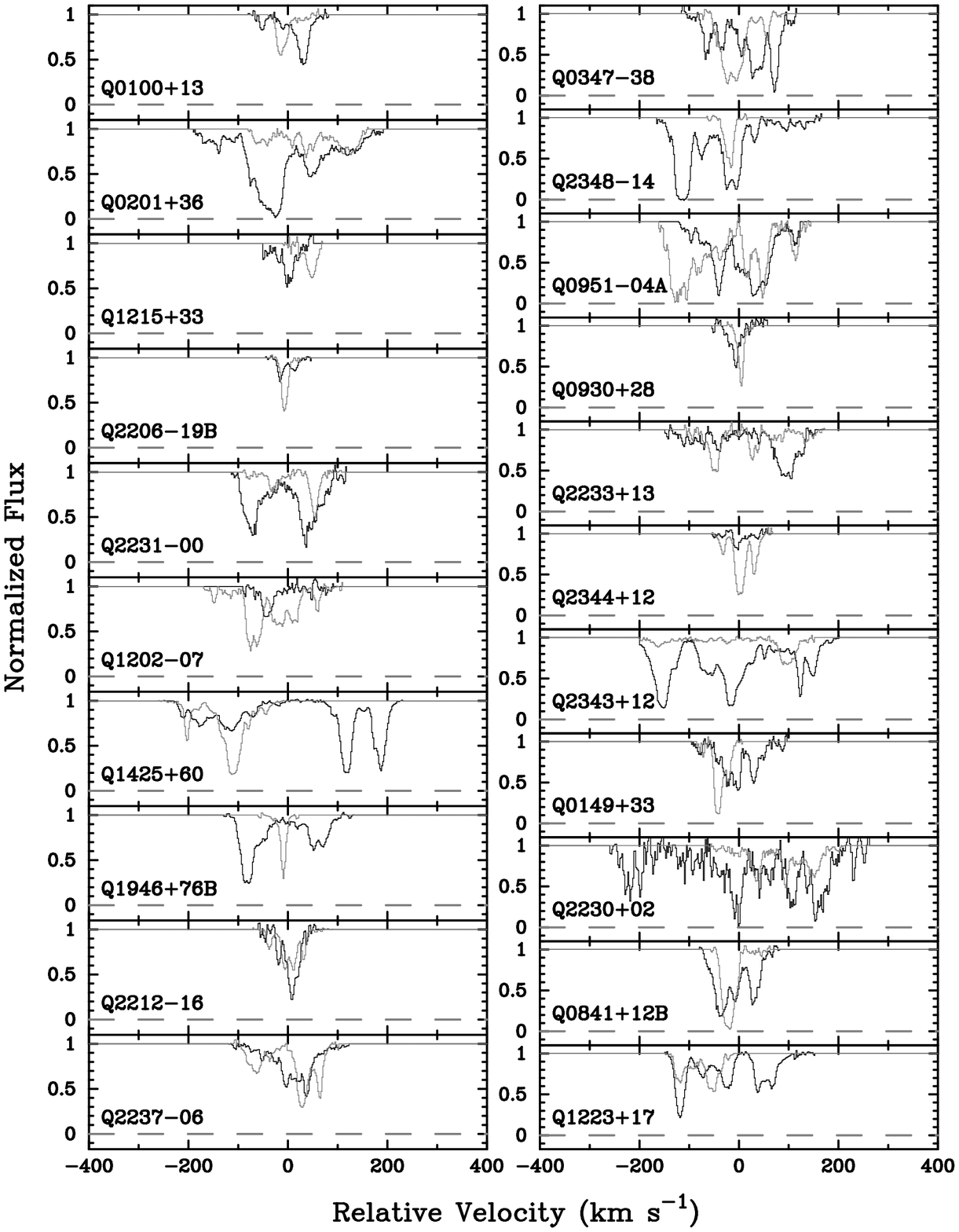}
\caption{Velocity profiles of
Si IV (dark lines) and low ion (grey lines) transitions for
all  damped {\lya} protogalaxies in
Table 1 in which HIRES Si IV and low ion profiles have been measured. Velocity scale
same as in Figure 1}
\label{SiIVvsLow}
\end{figure*}

\begin{figure*}[ht]
\includegraphics[height=5.8in, width=6.2in]{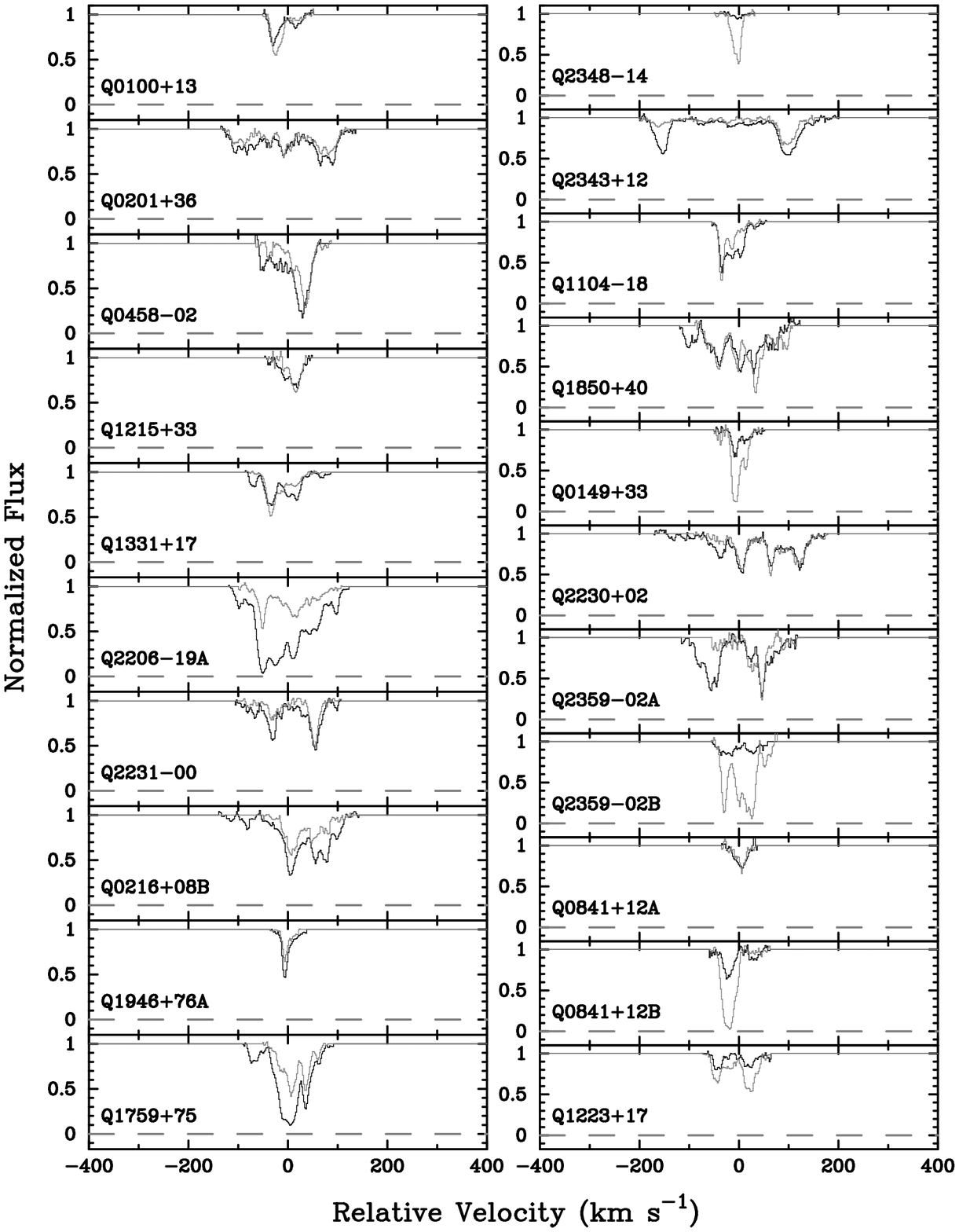}
\caption{Velocity profiles of Al III (dark lines) and low ion (grey lines)
transitions for all damped {\lya} protogalaxies in Table 1 in which HIRES Al III
and low ion proflies
have been
measured. Velocity scale same as in Figure 1}
\label{Al3vsLow}
\end{figure*}

\begin{figure*}[ht]
\includegraphics[height=5.8in, width=6.2in]{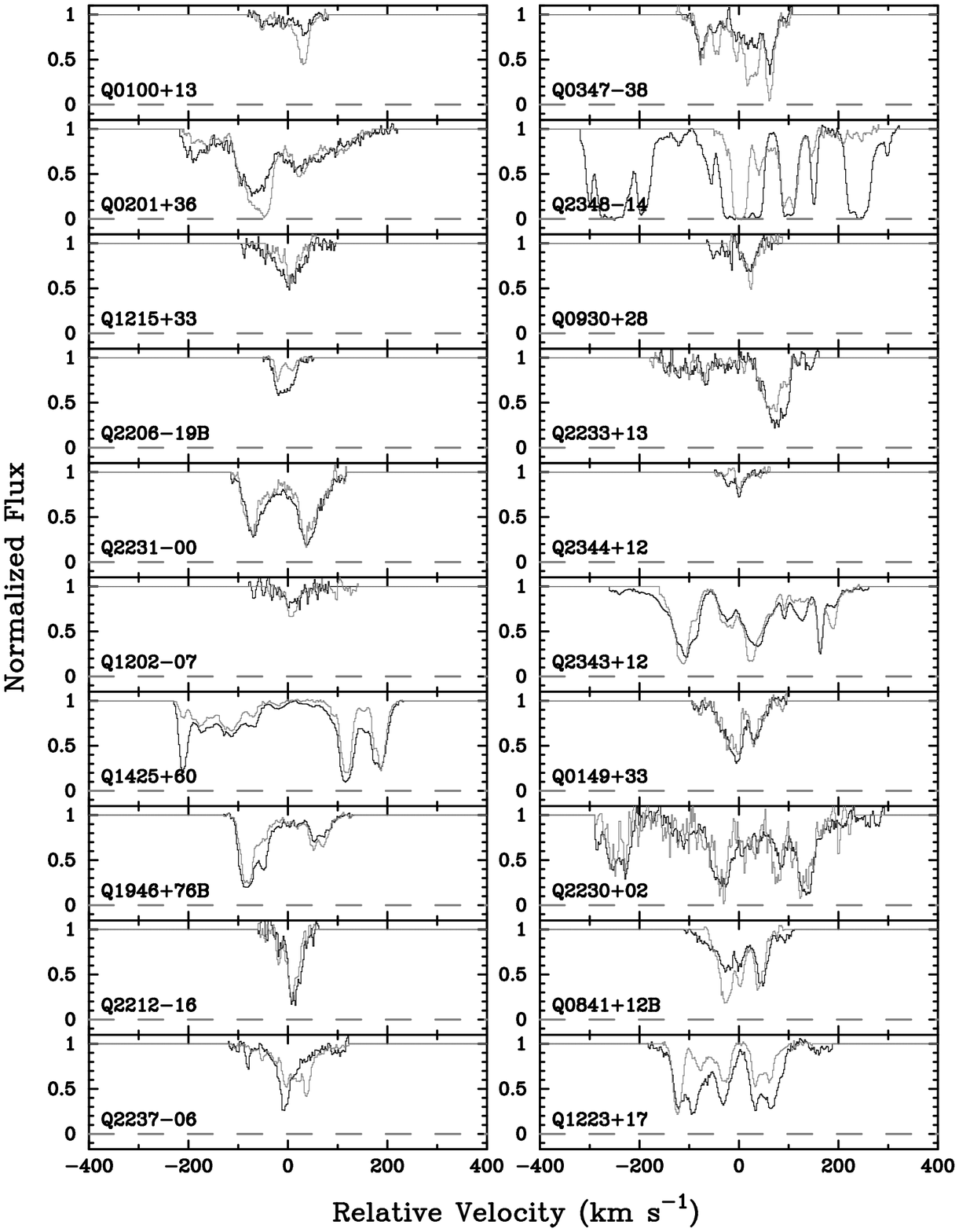}
\caption{Velocity profiles of C IV (dark lines) and Si IV (grey lines)
transitions for all damped {\lya} protogalaxies in Table 1 in which HIRES Si IV
and C IV profiles have been measured. Velocity scale same as in Figure 1.}
\label{CIVvsSiIV}
\end{figure*}

\begin{figure*}[ht]
\includegraphics[height=5.8in, width=6.2in]{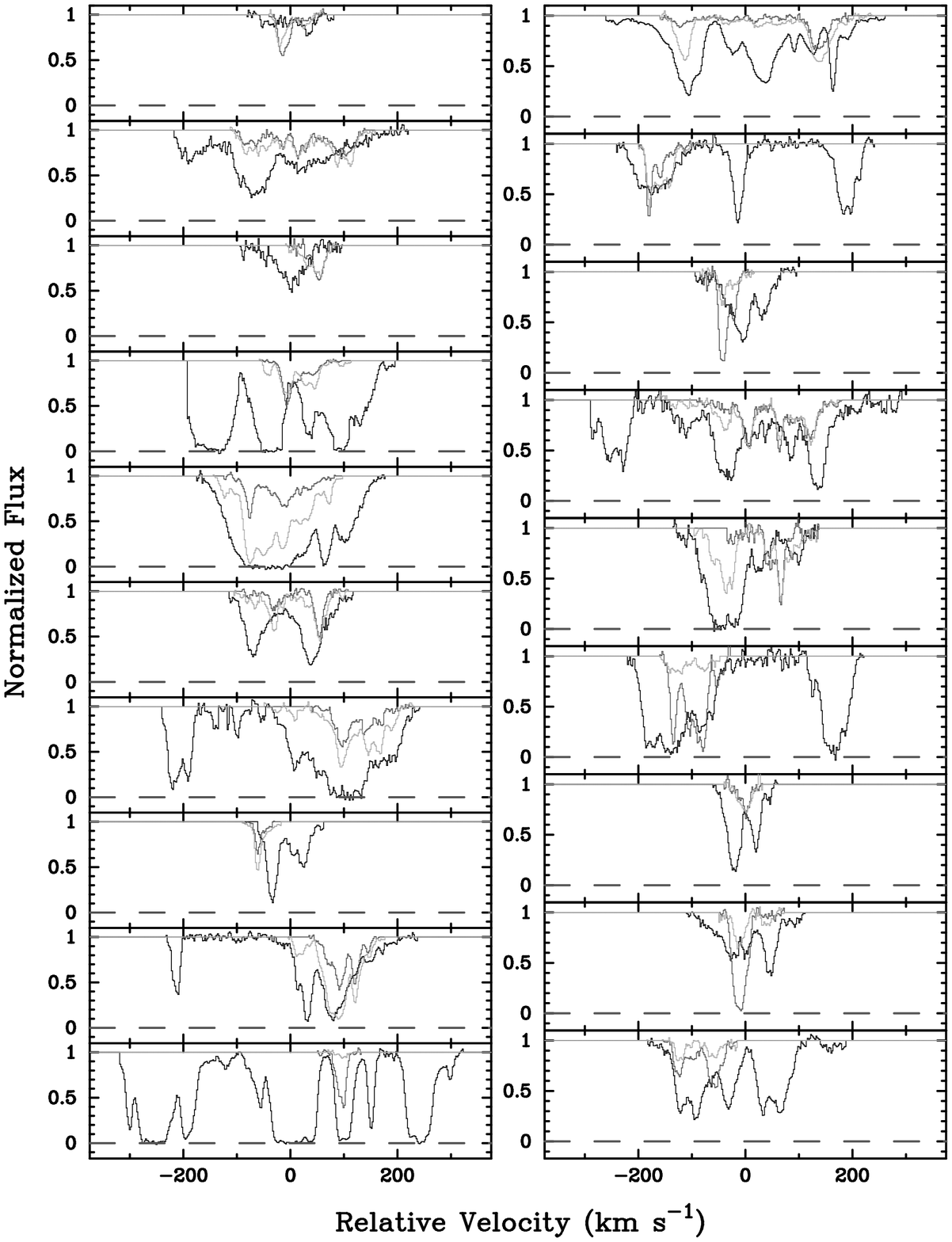}
\caption{Velocity profiles of  C IV (dark lines),
Al III (light lines), and
low ions (grey lines) transitions for all damped {\lya} protogalaxies
in Table 1 in which Al III, C IV, and low ion profiles have been measured.
Velocity scale same as in
Figure 1}
\label{CIVvsAl3vsLow}
\end{figure*}

The 
velocity profiles are shown in Figures ~\ref{CIVvsLow}$-$~\ref{CIVvsAl3vsLow}
which plot flux versus velocity, where the flux, $I(v)$, is
normalized to unit continuum. In Figures ~\ref{CIVvsLow}$-$~\ref{Al3vsLow} we compare
C IV, Si IV, and Al III  with the corresponding low ion profiles.
The dark curves depict the high and intermediate ions and the grey curves depict the
low ions. Figure~\ref{CIVvsSiIV} compares the C IV (dark curves) and
Si IV (grey  curves) profiles, and Figure~\ref{CIVvsAl3vsLow} compares C IV
(dark curves), Al III (light curves), and the low ions (grey curves).
In all cases, $v$ = 0 {\kms} corresponds to the redshifts in Table 1.
The low ion transitions were selected  on the basis of criteria outlined
in PW1 and PW2; i.e.,
for high signal-to-noise ratios and absence of saturation. The high ion transitions
were selected according to the same criteria, where possible. In
some cases the only transitions available exhibit saturation over sizable
velocity intervals.
The C IV profiles toward  Q0216$+$08B, Q1331$+$17, Q2206$-$19A, Q2348$-$14,
and Q2359$-$02A
contain significant regions in velocity space with strong saturation.
The same is true for
the Si IV profile toward  Q2348$-$14.
As emphasized in PW1, some of the statistical tests
used to compare the data with model predictions are sensitive to saturation;
in particular determination of test statistics measuring the profile asymmetries
can be inaccurate in the presence of saturation. However, the results
of statistical tests described below did not change qualitatively when
the saturated C IV profiles were excluded from the sample.
As a result
the saturated profiles shall be included in all subsequent analysis.

The velocity profiles leave the following impressions:

$\bullet$ In common with the low ion
profiles, the intermediate and high ion profiles exhibit a multi-component
structure comprising several narrow
components spanning a wide range of velocity intervals.

$\bullet$
The C IV and low ion profiles in Figure~\ref{CIVvsLow} are kinematically
disjoined. In many cases strong C IV absorption components
are at velocities where
low ion absorption is weak or absent. In other cases C IV absorption
is weak or absent at velocities where low ion absorption is strong.
In many cases the velocity centroids of the C IV and low ion profiles appear
to be significantly different.

$\bullet$
Despite their differences, the C IV and low ion velocity profiles overlap in
velocity space in such a way that the low ion profiles generally lie within
the high ion profiles.

$\bullet$
The low ion profiles also appear to be kinematically
disjoined from the Si IV profiles (see Figure~\ref{SiIVvsLow}).

$\bullet$
The Al III and low ion profiles appear to be strongly correlated (Figure~\ref{Al3vsLow}).

$\bullet$
The C IV and Si IV profiles appear to be strongly correlated (Figure~\ref{CIVvsSiIV}).

$\bullet$
In many cases C IV absorption occurs in regions of velocity space that are
free of Al III and low ion absorption
(Figure~\ref{CIVvsAl3vsLow}).

$\bullet$
In most cases the velocity widths of C IV and Si IV velocity profiles
exceed the velocity widths  of the Al III and low ion profiles.

$\bullet$
Although the low ion and high ion velocity profiles exhibit ``edge-leading''
asymmetries in which  the strongest velocity components are at the profile edges,
the high ions tend to exhibit peaks at both edges more often
than the low ions.

In the sections that follow, we introduce several tests to quantify
these phenomena.

\section{TEST STATISTICS}

\subsection{Distributions for single ions}

To describe the kinematics of the gas in a quantitative manner, we consider
test statistics that characterize the extent and shape of the velocity
profiles. We discuss distributions of such statistics drawn
from the profiles corresponding to given ions and then compare
them.
These
empirical distributions are used to
test theoretical models in  paper II.

In PW1 we
characterized
the extent and asymmetry of the velocity profiles
with four test statistics.
Briefly stated they are:

(1) {\dv}, the velocity interval enclosing the central
90$\%$ of the integrated optical depth,
$\int {\tau(v)}dv$, where $\tau(v)$ $\equiv$ ln[1/$I(v)$].

(2) $f_{mm}$, the ``mean-median statistic'' given by
$|v_{median}-v_{mean}|$/({\dv}/2), where $v_{median}$ and $v_{mean}$ are the median and mean
velocities of the profiles.

(3) $f_{edg}$, the ``edge statistic''
given by $|v_{peak}-v_{mean}|$/({\dv}/2),
where $v_{peak}$ is the velocity of the absorption component with peak optical depth.

(4) $f_{tpk}$, the  ``two-peak statistic'' given by \\ $\pm$$|v_{tpk}-v_{mean}|$/({\dv}/2), where
$v_{tpk}$ is the velocity of the component with second strongest peak optical-depth:   $f_{tpk}$
 is
negative if $v_{peak}$ $<$ $v_{tpk}$ $<$ {\dv} and positive if 0 $<$ $v_{tpk}$ $<$ $v_{pk}$.
If necessary
the optical depth profiles have been reflected in velocity space in such
a way that $v_{peak}$  is always located at $v$ $\le$ $v_{mean}$.

\begin{figure*}[ht]
\includegraphics[height=5.8in, width=6.2in]{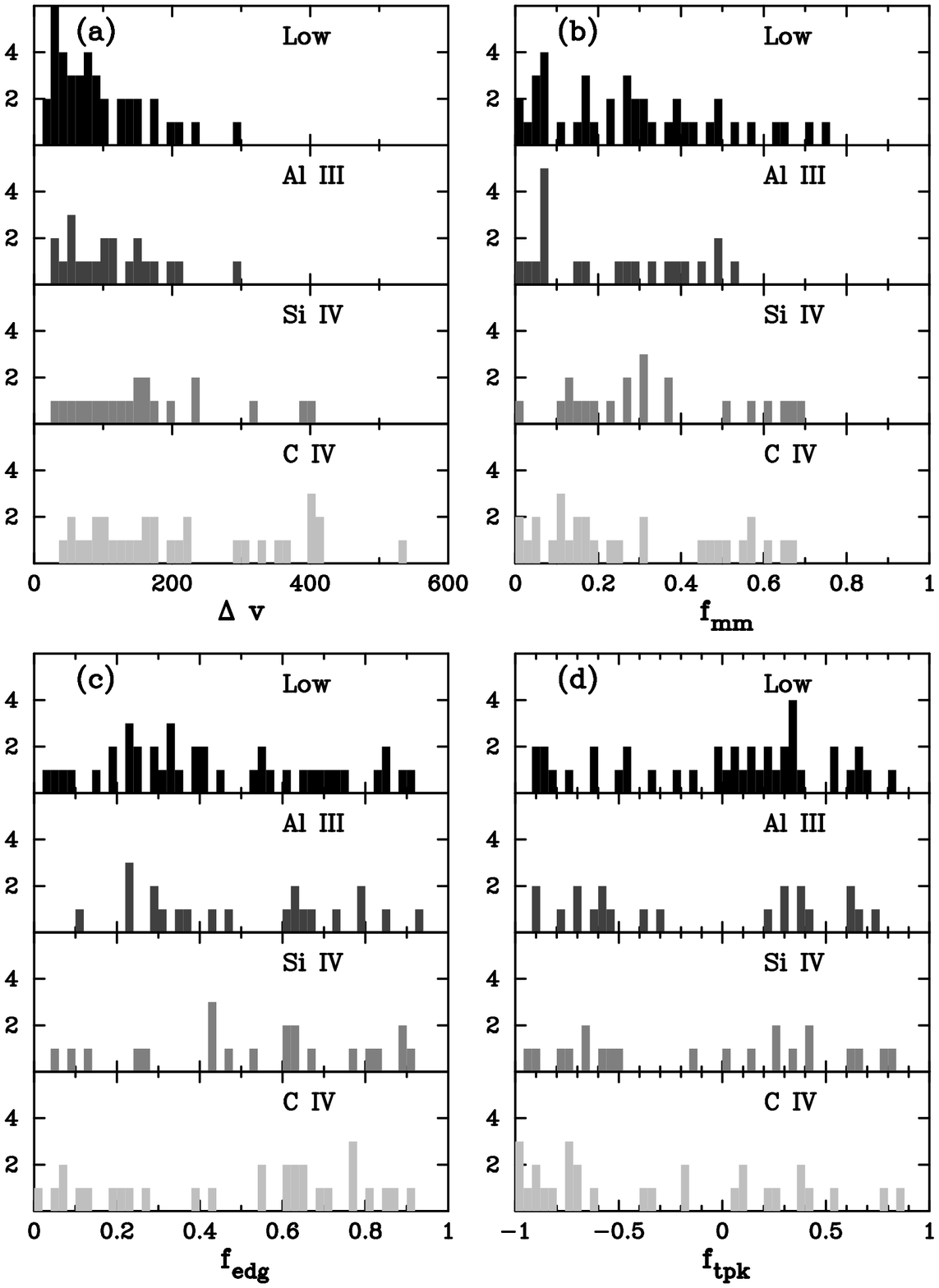}
\caption{ Empirical distributions  of test statistics (a) {\dv}, (b) $f_{mm}$,
(c) $f_{edg}$, and (d) $f_{tpk}$ for
 low ion, Al III, Si IV, and C IV velocity profiles.}
\label{origstat}
\end{figure*}

Distributions for each test statistic are illustrated in Figure~\ref{origstat}  for the low ion,
Al III, Si IV, and C IV transitions. In PW1 and PW2 we focused on the low ion distributions.
We used them to argue in favor of models in which the low ions were confined
to rapidly rotating disks (rotation speeds  $\approx$ 250 {\kms}) and to rule out
several other models. We return to this topic in paper II. Here
we
check the null hypothesis that low ion and high ion kinematics
stem from the same process. We do this by computing
${P_{\rm KS}}$, the Kolmogorov-Smirnov probability that each high ion
or intermediate-ion test statistic is
drawn from the same parent population as the corresponding low ion statistic.
Figure~\ref{origstat} shows that the null
hypothesis cannot be dismissed, with two exceptions. First,
the null hypothesis is highly unlikely when
comparing low ion versus C IV   distributions of {\dv} since $P_{\rm KS}$({\dv})= 0.002,
and when comparing low ion versus Si IV distribution of {\dv} since $P_{\rm KS}$({\dv}) = 0.021:
the high ions do not exhibit as much power at low {\dv} as the low ions,
and  more power than the low ions at {\dv} $>$ 200 {\kms}. Indeed, the low probabilities
suggest the low ions and high ions are in two distinct kinematic
subsystems.
Secondly, the
null hypothesis is unlikely in the case of low ion versus C IV
distributions of $f_{tpk}$ because $P_{\rm KS}$($f_{tpk}$) = 0.035;
i.e., the C IV profiles display more double
peaked profiles than the low ions. On the other hand $P_{\rm KS}$($f_{tpk}$)
= 0.23 in the case of low ion versus Si IV profiles.
Since
compatibility of  the Si IV and low ion distributions of
$f_{tpk}$ cannot not be ruled out, this is the only case in which the C IV and Si IV kinematics
differ. We suspect that statistics of small numbers may be affecting the
Si IV distribution,  a possibility we will test when larger
samples become available.

\subsection{Correlations Between Macroscopic Kinematic Properties  of Ion Pairs}
\label{sec_corr}

We now turn to correlations between kinematic properties of
gas in different ionization states. First we consider
distributions of
the following new
test statistic:

(5) $\delta v$ = $v_{mean}$(ion $a$) $-$
$v_{mean}$(ion $b$), i.e., the difference between the means of the velocity
profiles of ion $a$
and ion $b$.

\begin{figure*}[ht]
\includegraphics[height=5.8in, width=6.2in]{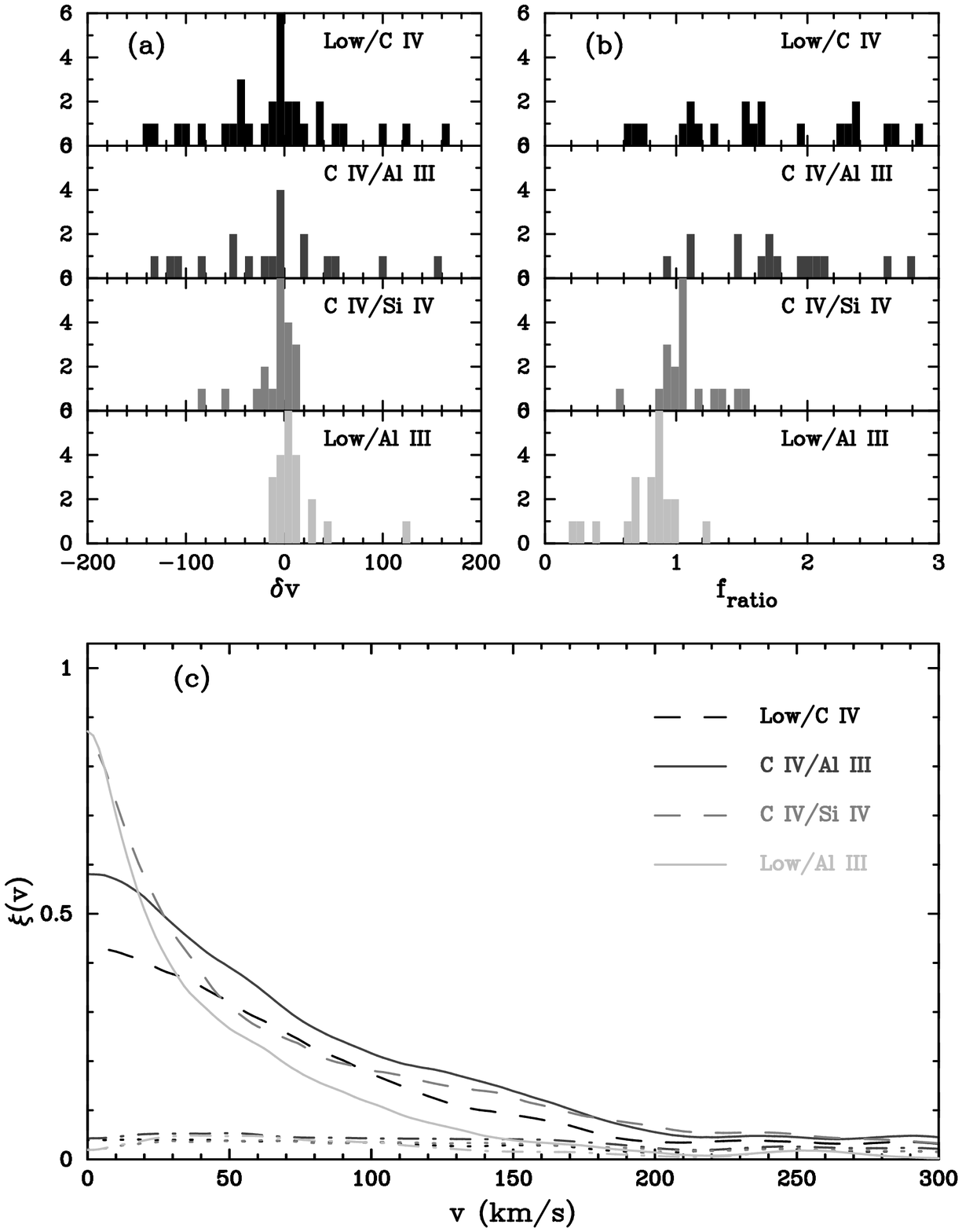}
\caption{ Empirical distributions of correlation
test statistics (a) {\dvm} and (b) $f_{ratio}$, and  cross-correlation
functions (c) $\xi_{ab}(v)$ for  low ion versus C IV (dark dashed curve), C IV versus
Al III (dark solid curve), C IV versus Si IV (light dashed curve), and low ion
versus Al III (light solid curve) velocity profiles.
Lower dashed curves are 1$\sigma$ errors.}
\label{figocorr}
\end{figure*}

\noindent Figure~\ref{figocorr}a shows the distribution of {\dvm} when $a$ and $b$ are 
(i) low ion
and C IV, (ii) C IV and Al III, (iii) C IV
and Si IV, and (iv) low ion and Al III. Not surprisingly, the distribution widths are narrower f
or
C IV versus Si IV and low ion versus Al III than for low ion versus C IV.
This just indicates that the high ion subsystem includes the high ions
C IV and Si IV, but not the intermediate Al III ions which
are mainly associated with the low ion subsystem.
The standard deviations for the 3 distributions are
given in Table 2. Although the dispersion of the low ion versus C IV
distribution is relatively high, $\sigma_{\delta}$ = 67 {\kms},
it is sufficiently low to place crucial restrictions on most models (see paper II).

\begin{table*}[h!!]
\begin{center}
\caption{DISPERSIONS of $\delta v$ DISTRIBUTIONS} \label{tab2}
\begin{tabular}{lccc}
\tableline
\tableline
&\multicolumn{3}{c}{}\\
\cline{2-4}
Dispersion & C IV versus Low
& C IV versus Si IV
& Al III versus Low \\
\tableline
$\sigma_{\delta v}$\tablenotemark{a} ({\kms}) & 67.0$\pm$8.2 & 23.2$\pm$3.4    & 30.0$\pm$4.2  \
cr
\end{tabular}
\end{center}
\tablenotetext{a}{One sigma error in {\sigdel} computed by assuming (n$-$1)s$^{2}/{\sigma^{2}}$
is distributed as  $\chi^{2}$(n$-$1) where n is the number of  degrees of freedom,
 $\chi^{2}$(n$-$1) is $\chi^{2}$ for n$-1$ degrees of freedom,  and
where s$^{2}$ and $\sigma^{2}$ are the measured and true
variance of {\sigdel} (Frodesen {\etal} 1979) }
\end{table*}

\begin{table*}
\begin{center}
\caption{CORRELATIONS BETWEEN IONIC VELOCITY INTERVALS} \label{tab3}
\begin{tabular}{lcccc}
&\multicolumn{4}{c}{Kendall $\tau$}\\
\cline{2-5}
ION & C IV
& Si IV
& Al III
& Low  \\
\tableline
C IV & ...& 0.80$\pm$0.16   &0.57$\pm$0.16 & 0.33$\pm$0.16   \cr
Si IV   & ... & ...    & 0.76$\pm$0.22& 0.52$\pm$0.15   \cr
Al III & ... & ...& ....  & 0.73$\pm$0.16   \cr
\end{tabular}
\end{center}
\end{table*}

\begin{figure*}[h]
\includegraphics[height=5.8in, width=6.2in]{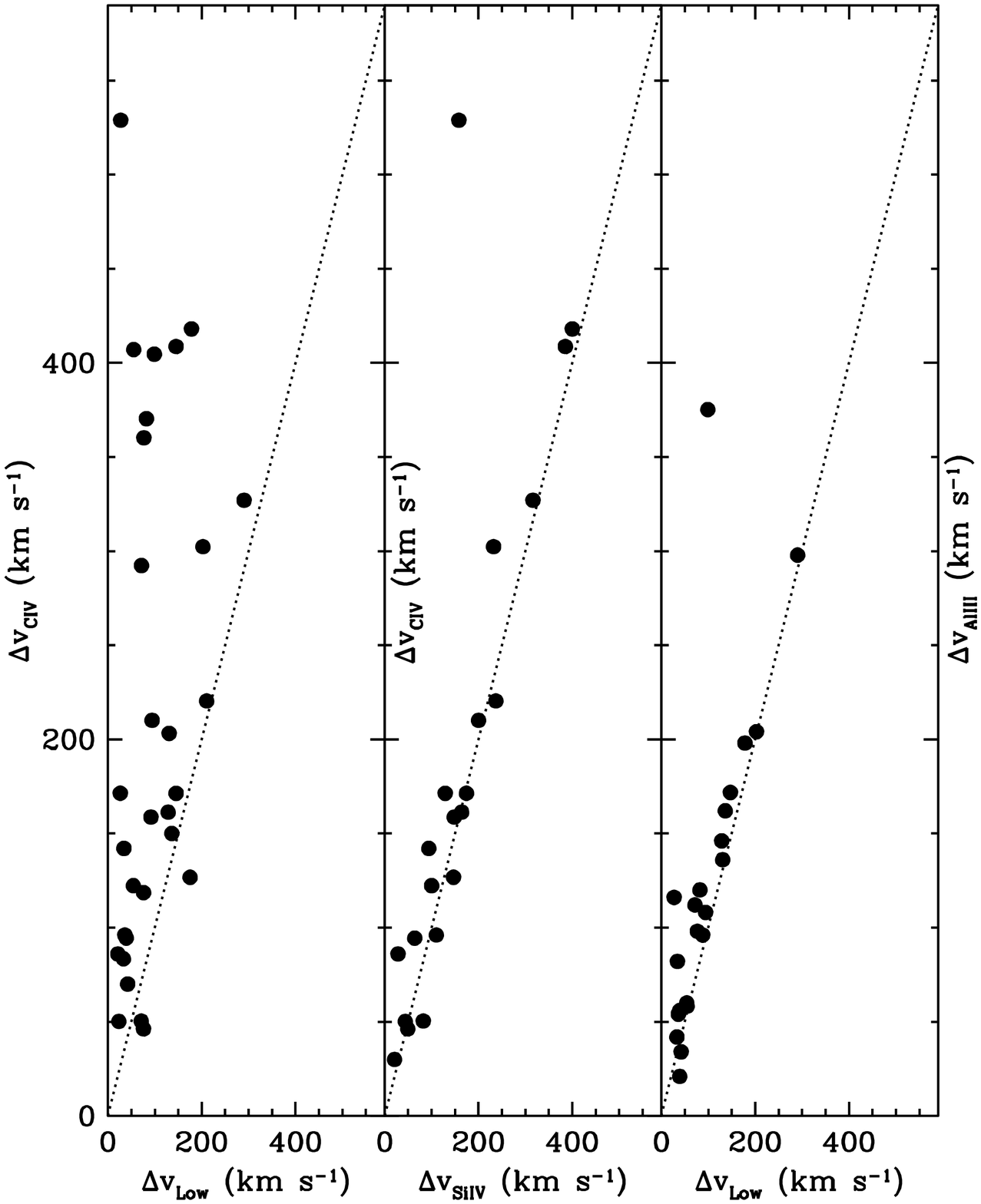}
\caption{ Plots of {\dv} for the ion pairs (a) C IV versus Low,
(b) C IV versus Si IV, and (c) Al III versus Low}
\label{delvpairs}
\end{figure*}

Next we consider correlations between the {\dv} of various ion pairs.
In Figure~\ref{delvpairs}a we plot {\dciv}, the {\dv}
inferred from the 
C IV profiles versus {\dlow}, the {\dv} inferred from the low ion profiles.
In Figures ~\ref{delvpairs}b and ~\ref{delvpairs}c we do the same for
{\dciv} versus {\dsiiv}, the velocity width of Si IV, and {\daliii}, the velocity
width of Al III versus {\dlow}.
The figure shows {\dciv} to be correlated with
{\dsiiv} and {\daliii} with {\dlow}.
This is demonstrated in Table 3  showing
the Kendall $\tau$ correlation coefficients. The only cases
with statistically significant {$\tau$}
(i.e., at the 5$\sigma$ level) are
for the C IV versus Si IV correlation
and the Al III versus low ion
correlation. We conclude
that statistically significant correlations exist between {\dciv} and {\dsiiv},
and between {\daliii} and {\dlow}.
Evidence for such correlations
within the
remaining ion pairs is marginal.
On the other hand Figure~\ref{delvpairs}a shows
that for all but 3 objects, {\dciv} $\ge$ {\dlow},
and even the three exceptions nearly satisfy this relation.
In other words the low ion
velocity width acts as a ``floor'' to the high ion velocity widths. While
not a correlation, this is an important systematic effect indicating
that the high ion and low ion systems are interrelated in a way that
must be accounted for by models of the ionized gas.

The relationship between the velocity widths
is quantitatively described by the test statistic:

(6) $f_{ratio} \ \equiv$ \ ${\Delta v}_{a}$/${\Delta v}_{b}$, the ratio of
the velocity widths of ion $a$ to those of ion $b$.

\noindent The corresponding distributions are shown in Figure~\ref{figocorr}b.
Notice how the $f_{ratio}$ for low versus C IV  and C IV versus Al III are
uniformly distributed   at 0.7 $\le$ $f_{Ratio}$ $\le$ 3, while the
C IV versus Si IV and low versus Al III distributions exhibit significant peaks
at $f_{ratio}$ = 1. This is further
evidence for an association between Al III and low ions,
and between C IV and Si IV.

\begin{figure*}[h]
\includegraphics[height=5.8in, width=6.2in,angle=-90]{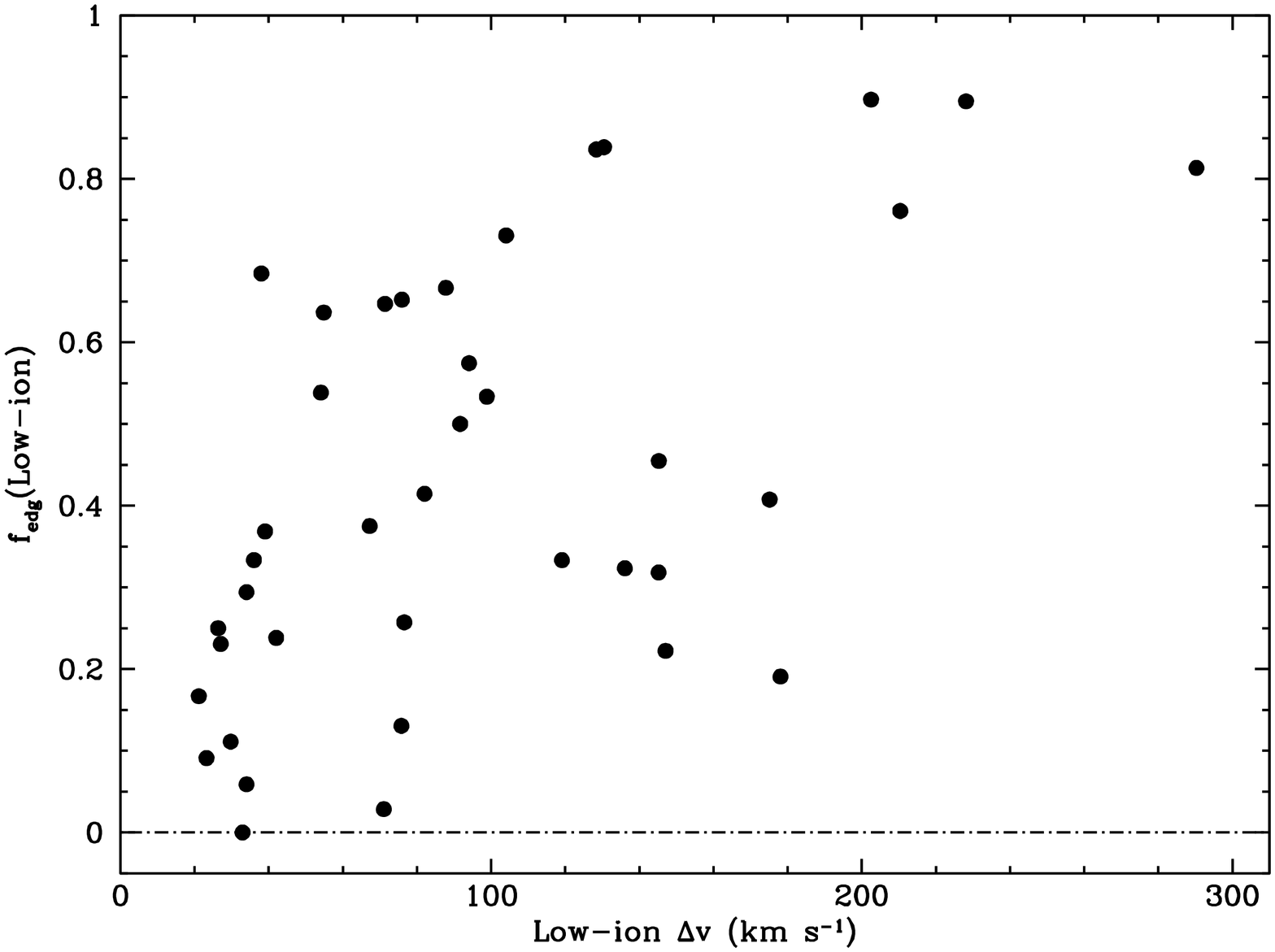}
\caption{Plot of $f_{edg}$ versus {\dv} for low ion profiles in
Table 1}
\label{fedgdv}
\end{figure*}

We searched for evidence that other test statistics were
correlated between the different ion pairs. None was found, with the possible exception of
correlations between the Al III and low ion $f_{edg}$ which exhibit Kendall
$\tau$ coefficients at
the 3.0$\sigma$ level. Correlations between between the C IV and Si IV
$f_{tpk}$ were found at the 2.5$\sigma$ level.
We also checked for correlations between
pairs of test statistics corresponding to a given ion.
\cite{ledx98} reported evidence for a low ion correlation
between $f_{edg}$ and {\dv} out to {\dv} = 150 {\kms}, but
found no such correlation at
{\dv} $>$ 150 {\kms}.
They claimed the ``break'' at 150 {\kms}  argued against
the presence of damped {\lya} systems comprised of
disks with rotation speeds greater than 150 {\kms}. It is
difficult to assess the validity of their claim since no quantitative
estimate was given for the significance of the correlation.
With our larger data set (39 versus their 26 systems)
we find evidence for such a correlation
at the 3.6$\sigma$ level, but no convincing evidence for a ``break''
at {\dv} $\approx$ 150 {\kms}. In fact when  systems with
{\dv} $>$ 150 {\kms} were eliminated, the significance level of the correlation
{\em decreased} to 2.8$\sigma$. This argues against the presence
of such a ``break'' and is consistent with a correlation extending to
{\dv}
$\approx$ 250 {\kms}. As a result disks with
rotation speeds exceeding 150 {\kms} cannot be ruled out. In any case, the  correlation
between $f_{edg}$ and {\dv} is tentative (see Figure~\ref{fedgdv}),
and needs to be tested with larger data
sets.

\subsection{Cross Correlation Functions}
\label{sec_cross}

We next investigate whether the detailed component structures
exhibited by the velocity profiles of various ions are correlated. In Figure~\ref{figocorr}c we
plot
$\xi_{ab}(v)$,
the cross-correlation functions between ions $a$ and $b$
versus lag velocity $v$, and the 1$\sigma$ errors, $\sigma_{\xi_{ab}}(v)$.
The $\xi_{ab}(v)$ are given by

\begin{equation}
{\xi_{ab}(v)} = {1 \over N} \sum_{k=1}^{N}{1 \over {\sqrt{\xi_{aa}^{k}(0)\xi_{bb}^{k}(0)}}} \  {
\sum_{i=1}^{n_{k}}}{{[\tau_{ak}(u_{i})]
[\tau_{bk}(u_{i}-v)]}} \cmma
\label{crosscor}
\end{equation}

\noindent where  the first sum is over
$N$  profiles, and the second
is over $n_{k}$
velocity channels, $u_{i}$, spanning the $k^{th}$ absorption
profile. The quantity $\tau_{ak}(v)$ is the
optical-depth profile of ion $a$
for the $k^{th}$ profile. The quantity, $\xi_{aa}^{k}(0)$, is the autocorrelation
function for ion $a$ in the $k^{th}$ profile at $v$ = 0, and  is defined as

\begin{equation}
{\xi_{aa}^{k}(0)} =  {\sum_{i=1}^{n_{k}}}{{[\tau_{ak}(u_{i})]^{2}
}} \perd
\end{equation}

\noindent As a result the cross-correlation functions are normalized  so that
perfectly correlated ions yield $\xi_{ab}(v)$ = 1 at zero lag velocity.
The errors are dominated  by cosmic variance and are computed by
a ``bootstrap'' method \citep{ds86}.

Comparison between the cross-correlation functions
and the  errors demonstrates that
in some cases the measured $\xi_{ab}(v)$ are statistically
significant out to $v$ $\sim$ 150 {\kms}.
The strongest correlations shown are between the low ion and Al III
profiles and between the C IV and Si IV profiles.
This confirms the rather accurate one-to-one alignment between the velocity components
comprising these ion pairs
(see Figures~\ref{Al3vsLow} and ~\ref{CIVvsSiIV}).
The  20 {\kms} half-width of $\xi_{ab}$ for Al III versus low 
roughly equals the half widths of the wider components in Figure~\ref{Al3vsLow},
which supports the idea of common velocity components for Al III and low ions.
The 30 {\kms} half width  of $\xi_{ab}$ for C IV versus Si IV
argues for common component structures for these two ions.

By contrast
the $\xi_{ab}(v)$ for the C IV versus low ions or C IV versus Al III
exhibit significantly lower, but still statistically significant, values at $v$ = 0
(the differences exceed 6$\sigma$).
The lower $\xi_{ab}(0)$ stems from
misalignment between the low ion (or Al III) velocity components with
the C IV components in many of the profiles.
In this case the correlation amplitude arises from overlap in velocity space between
the ion pairs rather than
one-to-one alignment between individual components.
This interpretation is supported by the $\sim$
80 {\kms}
half widths of $\xi_{ab}$  for the C IV versus Al III or C IV versus low profiles.
These half-widths more closely
resemble the coherence scales of contiguous multiple components in the C IV profiles
than the half widths of individual components (see Figure~\ref{CIVvsLow}).

\section{SUMMARY AND CONCLUDING REMARKS}

Using high-resolution absorption spectra obtained with HIRES,
the Echelle spectrograph on the Keck I telescope  \citep{vog94},
we have probed the kinematics of ionized gas in a sample of 35
high-redshift damped
{\lya} protogalaxies. Specifically, we obtained velocity profiles of the high ions
C IV and Si IV     and the intermediate ion Al III. We studied the
kinematic state of the gas by constructing empirical test statistics which
characterize (a) the widths and symmetry properties of high ion,
intermediate-ion, and previously obtained low ion velocity profiles,
and (b) correlations between the kinematic properties of various ionic pairs.

What have we learned that is new? We answer this question
by discussing model-independent
conclusions inferred from the data:

(1) The damped {\lya} absorbers comprise distinct kinematic subsystems:
a low ion subsystem in which low ions such as Al II    are
physically associated with intermediate ions such as Al III    ,
and a high ion subsystem consisting of ions such as C IV     and
Si IV. This is indicated by  the similarity between the
C IV and Si IV
velocity profiles and between the low ion and Al III velocity profiles,
and by the misalignment of the velocity components
comprising the C IV and low ion profiles.

(2) The low ion and high ion kinematic subsystems are related despite
misalignment of their velocity components. This follows from
the detection of a statistically significant C IV versus low ion
cross-correlation function which exhibits lower amplitude and
a wider half-width
than either the low ion versus Al III, or C IV versus Si IV cross correlation
functions. Whereas the high cross-correlation amplitudes of the latter two
ionic pairs arise
from the coincidence between corresponding velocity components,
the lower amplitude of the C IV versus low ion cross-correlation 
function is due to a general overlap in velocity space between
the line profiles. In any case the relation between
low ion and high ion subsystems is
further indicated by a systematic
effect in which
{\dciv} $\ge$ {\dlow} in 29 out of 32 profiles.

(3)  The difference between the mean velocities of the C IV and Si IV velocity profiles
or between the low ion and Al III profiles exhibits distributions with
dispersions  equaling 23 and 30 {\kms} respectively.
By contrast the low ion versus C IV distribution
has a significantly wider dispersion of 67 {\kms}.
This is more evidence for distinct kinematic
subsystems.

(4) The absence of intermediate ions and low ions from the
high ion subsystem indicates the latter is optically
thin at the Lyman limit. The lack of mixed ionization
states distinguishes the high ion gas
in damped {\lya} protogalaxies from high ion gas
in the ISM \citep{sav93}, high-redshift
Lyman-limit systems \citep{pro99b,pro00}, and $z$ $\sim$ 0.7 Mg II
selected absorbers \citep{church99} where the velocity
profiles of the low ions and intermediate-ions
resemble those of the high ions.

These results have rather general implications.
First, kinematic subsystems placed in the same potential well
generate velocity profiles that  tend to overlap in velocity space.
In cases where the velocity field of the
neutral gas is constrained to fewer degrees of freedom than the ionized gas
the resulting profiles
will be characterized by
{\dciv} $\ge$ {\dlow}. Therefore, we shall test the hypothesis
that both kinematic subsystems are subjected to gravitational
forces arising from the same mass distribution
in paper II.
Second,
the absence of ionized gas optically thick at the Lyman limit
implies a distinction between the damped {\lya} systems
on the one hand
and
the Lyman limit systems and Mg II absorbers on the other.
This is potentially significant because most models assume
that both classes of absorber arise in the same physical object.
Finally, in paper II  we use the
empirical distributions of test statistics as well as the
empirical C IV versus low cross-correlation function
to constrain semi-analytic
models of galaxy formation.  Specifically  we focus on models
in which ionized gas in virialized dark-matter
halos falls onto centrifugally supported neutral hydrogen disks.

\acknowledgements

We wish to thank
Eric Gawiser and David Tytler for valuable discussions.
We also thank Wal Sargent for generously giving us data prior
to publication. AMW was partially
supported by NSF grant AST0071257 and JXP acknowleges support from a Carnegie
postdoctoral fellowship.

\end{document}